# Topological valley transport under long-range deformations

Zhixia Xu[1, 2, *], Xianghong Kong[2, 3], Robert J. Davis[2], Dia'aaldin Bisharat[2],
Yun Zhou[4], Xiaoxing Yin[1, †], Daniel F. Sievenpiper[2, ‡]

[1]State Key Laboratory of Millimeter Waves, Southeast University, Nanjing 210096, China

[2]Electrical and Computer Engineering Department, University of California, San Diego, California 92093, USA

[3]Department of Electronic Engineering, Shanghai Jiao Tong University, Shanghai 200240, China

[4]Department of Mechanical and Aerospace Engineering, University of California, San Diego, California 92093, USA

Corresponding Author:

[*]Zhixia Xu: zxxucn@seu.edu.cn

[†]Xiaoxing Yin: xxyin@seu.edu.cn

[‡]Daniel F. Sievenpiper: dsievenpiper@eng.ucsd.edu

**Abstract:**
Edge states protected by bulk topology of photonic crystals show robustness against short-range disorder, making robust information transfer possible. Here, topological photonic crystals under long-range deformations are investigated. Vertices of each regular hexagon in a honeycomb crystalline structure are shifted randomly to establish a deformed system. By increasing the degree of random deformations, a transition from an ordered system to an amorphous system are investigated, where the close of topological bandgap is clearly shown. We further present comprehensive investigations into excitation methods of the proposed deformed system. Due to the lack of strict periodicity, excitation of topological edge modes becomes difficult. Chiral and linearly polarized sources as two different methods are investigated respectively. It is found that chiral sources are sensitive and rely on the ordered lattice. Even a weak long-range deformation can bring fluctuations to transmission. We further designed and fabricated metal-dielectric-metal sandwich-like samples working in the microwave band. Using linearly polarized source, we detected the existence of topological transport in the deformed system. This work investigates excitation and robustness of bulk topology against long-range deformations and may open the way for exploiting topological properties of materials with a deformed lattice.

Systems with spatial order are the predominate topic in physical science where every individual unit cell behaves the same. Much of this is motivated by the simplicity of the analysis, as the behavior of waves interacting in a system can be deduced elegantly from rigorous formulae once properties of the unit cell are known. Imperfections are usually undesirable because random scattering is usually unpredictable and may deteriorate the performance of an ordered system. However, scattering in systems can be regarded as an elastic process, information is not lost and random scattering is reversible [1]. As an example, photonic bandgaps, which widely exist in long-range ordered crystalline structures [2,3], can also be found in quasicrystals and amorphous structures without any translational symmetry [4–6]. To some extent, disorder opens an extra degree of statistical freedom to analyze the performance of systems, leading to potential applications based on artificial structures [7–10].

In particular, topology links order and disorder, attracting intense research interest recently [11–14]. Topological invariants defined within the Brillion zone of ordered structures divides photonic bandgaps into trivial and non-trivial bandgaps. When some symmetry of a system is broken, corresponding degeneracies vanish and non-trivial bandgaps will appear where edge states under topological protection are immune to disorder. Gyroscopes and gyromagnetic materials have been used to break time-reversal symmetry to realize spin Hall effects from mechanical to photonic systems [15–19]. Breaking the geometric symmetry also helps establish non-trivial bandgaps. For example, reducing crystalline symmetry can generate pseudospin modes [20]. Photonic crystals with C3 symmetry featuring inversion asymmetry can support topological valley transport [21–25]. Using chiral waveguides or perturbations in the cylindrical structures breaks z-symmetry and opens a nontrivial bandgap [26,27]. Within most of these structures, a long-range ordered lattice is required to be preserved with translational symmetry where the unit cell is repeated periodically over the entire space.

Robustness against defects is an interesting property of topological materials. Most studies add a short-range disorder to observe robust topological edge modes, but when the disorder of the whole structure increases, topological states will vanish [28]. Thus, it is important to investigate how much spatial order is necessary to hold topological states. The Bott index has been proposed as a substitute for the Chern number [29–31]. According to the Bott index, topological surface states at interfaces between free space and bulk of quasicrystals or amorphous systems have been studied [32–34]. Another emerging topic is that of topological edge modes existing at the interface of two different amorphous materials. Recently, amorphous systems with broken time-reversal symmetry have been reported with unidirectional edge modes along the interface between two different amorphous bulks [35,36]. It is natural, therefore, to further investigate topological transport between different deformed structures without breaking time-reversal symmetry. In this work, we randomly deform a honeycomb lattice. Robust valley transport is investigated. Photonic density of state (PDOS), as a statistical parameter, is calculated to evaluate the existence of photonic bandgap. With increasing degree of deformation, the valley bandgap becomes narrower until it eventually disappears. We further designed and fabricated two samples with straight and triangular interfaces respectively based on the 3D printing technique, and measured topological transport within the bandgap.

The original structure is a periodic honeycomb lattice, as shown in Fig. 1(a). Dielectric rods with two different radii are placed on vertices with C3 symmetry, opening a Dirac point at K/K' with the valley topological property [21–24,37–40]. We assume the height of the dielectric rods is infinite to simplify the model to a 2D structure in the x-y plane.

We further deform the whole lattice to investigate topological transport with long-range disorder. Inspired by the nature of foam that every node is shared by three bubbles, we propose the deformed systems as shown in Fig. 1(b), where vertices are shifted randomly. The region of deformation marked in the inset ensures each node still connects three convex hexagons, which is the most common and stable topology of a bubble cluster statistically [41].

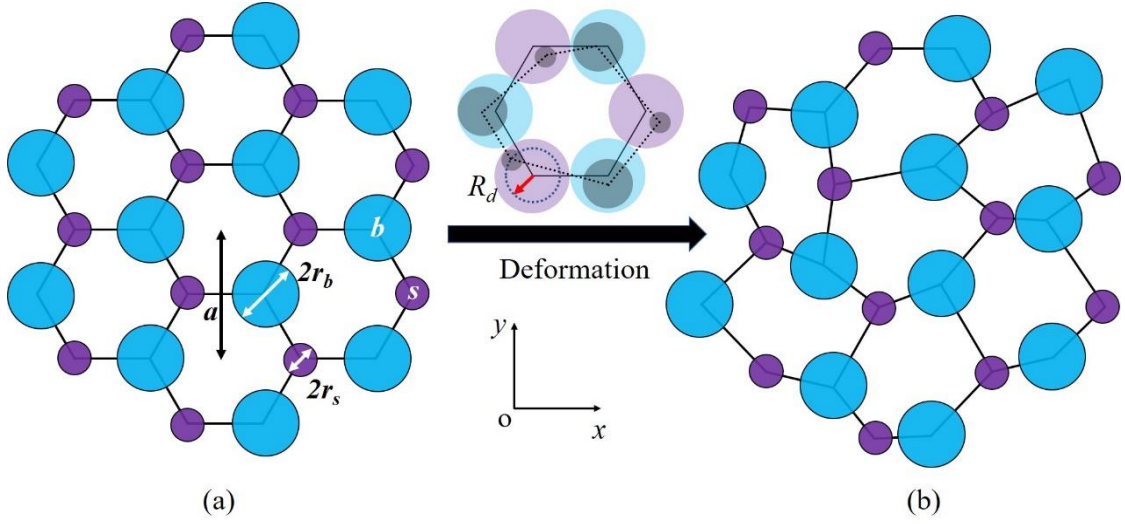

Fig. 1. (a) Honeycomb lattice with C3 symmetry whose geometric parameters are listed as: side length of hexagon lattice $L = 10$ mm; radius of rods $r_b = 0.4L$, $r_s = 0.2L$; lattice length is $a$, and the refractive index of all dielectric rods is set around 1.7. (b) Randomly deformed lattice: the center of every rod is random located in a circle centered in the original position whose radius $R_d$ can be regarded as the degree of deformation. The inset shows a situation where $R_d = L/2 - r_i$, $i = b$ or $s$, standing for rods with two different size.

Here, we consider TM modes with electric field perpendicular to the structure plane (x-y plane). We first present simulations of the original ordered structure without any deformation. Fig. 2(a) shows the simulated band diagram of a hexagonal unit cell with periodic boundary conditions. The dotted lines correspond to the honeycomb lattice with C6 symmetry where all rods are exactly the same size ($r_b = r_s = 0.4L$). Initially, there exists a degeneracy at K/K'. We then arrange two different rods in turn ($r_b = 0.4L$, $r_s = 0.2L$) to reduce the C6 symmetry to the C3 symmetry, thereby realizing a topological valley bandgap. Fig. 2(b) shows the phase distribution at valley states carrying intrinsic orbital angular momentum (OAM), counterclockwise or clockwise. Using a 1×12-size superlattice with periodic boundary conditions along the y-axis and electric boundary conditions along the x-axis, we can obtain the dispersion diagram of the edge mode supported by the small-rod interface shown in Fig. 2(c) and corresponding electric field distribution shown in Fig. 2(d), where energy is bound at the interface, propagating along the *y*-axis and decaying along the *x*-axis.

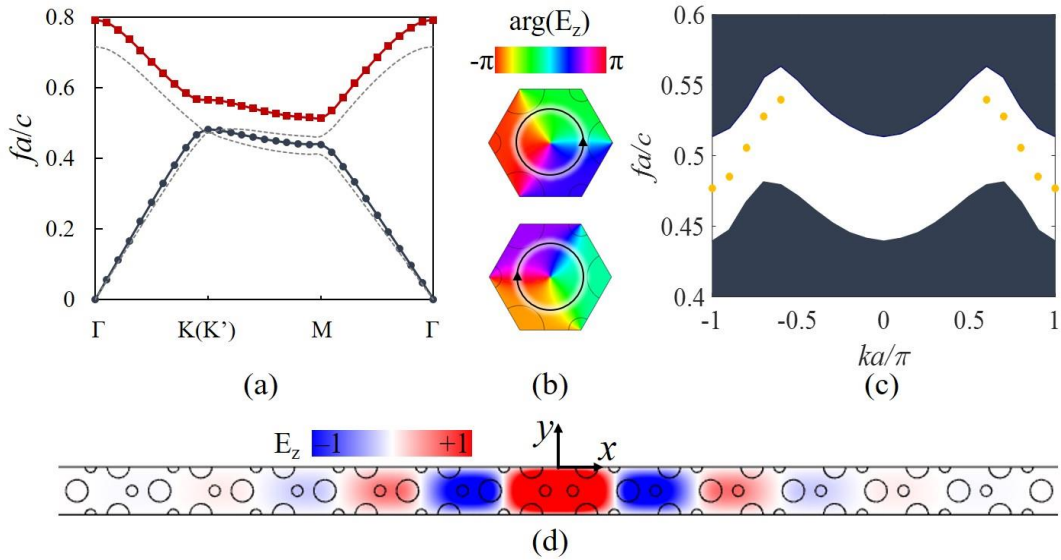

Fig. 2. (a) The opening of a valley bandgap. (b) Intrinsic OAM modes and phase distributions of $E_z$ at valley points. (c) Valley-dependent edge mode. (d) Edge mode supported by the interface.

When random deformations are added, translational symmetry of crystalline structures is no longer valid. As shown in the inset of Fig. 3, we set periodic boundary conditions around the perimeter as an approximation to suppose that the structure is infinite and further obtain eigenvalues. Fig. 3(a) shows eigenvalues of the superlattice under investigation, the horizontal axis shows the index of eigenmodes and the vertical axis represents the dimensionless frequency. We can observe a bandgap around 0.5 with several sparse points marked as red in the bandgap. A typical eigenstate within the bandgap is shown in Fig. 3(b), where most of the energy is distributed around boundaries of the superlattice. These eigen states can be regarded as errors brought on by the periodic boundary approximation. Fig. 3(c) shows a bulk state out of the bandgap, in which energy permeates into the whole superlattice.

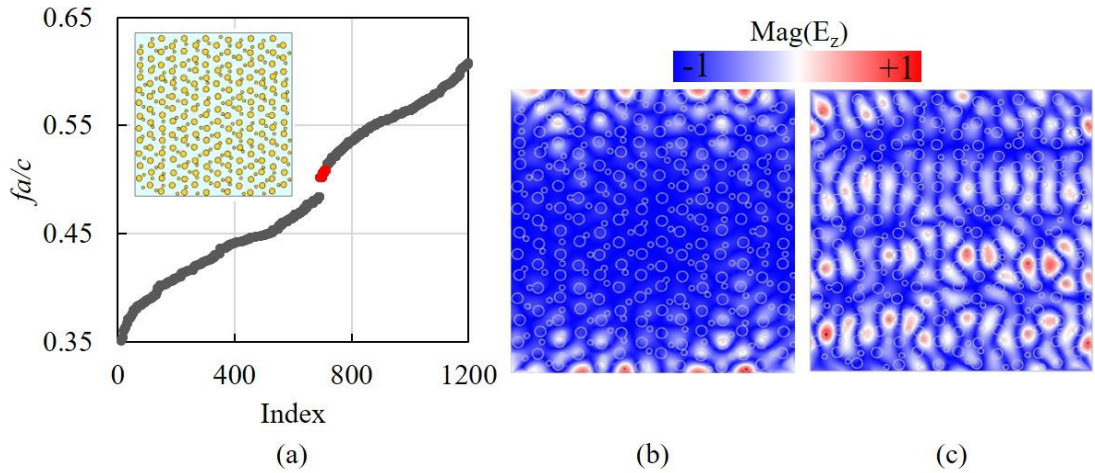

Fig. 3. (a) Distribution of eigenstates where the inset shows the superlattice with periodic boundary conditions. The degree of deformation is defined as $R_d = L/2 - r_i$. (b) An eigenstate within the bandgap. (c) An eigenstate out of the bandgap.

Based on distributions of eigenvalues, PDOS of these superlattices can be further calculated, as shown in Fig. 4. We investigate PDOS of four different superlattices undergoing different degrees of deformation defined by $R_d$. We coded the deformed pattern using MATLAB, then transferred it into COMSOL to establish models. Thus, there may exist some geometrical error during this process but it would not influence the conclusion. With the increasing degree of deformation to almost an amorphous distribution, the valley bandgap becomes narrower and it eventually disappears. The PDOS approach is a valid method to find bandgaps of non-periodic structures, such as quasi-crystals [31,32,34] and amorphous structures [10,33,35], where thousands of eigenvalues of a superlattice need to be calculated to detect bandgaps with enough accuracy. Thus, PDOS is a statistical parameter to analyze the deformed structure. We also calculate corresponding Fourier spectra of the superlattices. As the degree of deformation increases, the number of energy peaks in the reciprocal space decreases. The number of hexagonally-distributed points in the Fourier spectra shows the level of similarity between a deformed structure and the undeformed periodic lattice in real space. The greater the deformation, the fewer maximum points that can be found in the Fourier spectra. If hexagonally-distributed points disappear in the Fourier spectra, the structure can be regarded as an amorphous material whose bandgap disappears as well. In the presented examples, photonic bandgaps are clearly recognizable in Fig. 4 (a, b), become weak in Fig. 4(c), and finally disappear in Fig. 4(d). These identifiable points in the reciprocal space contain the average symmetry information of the lattice, and could be regarded as a limitation to judge the reservation of topological properties when materials undergo deformation.

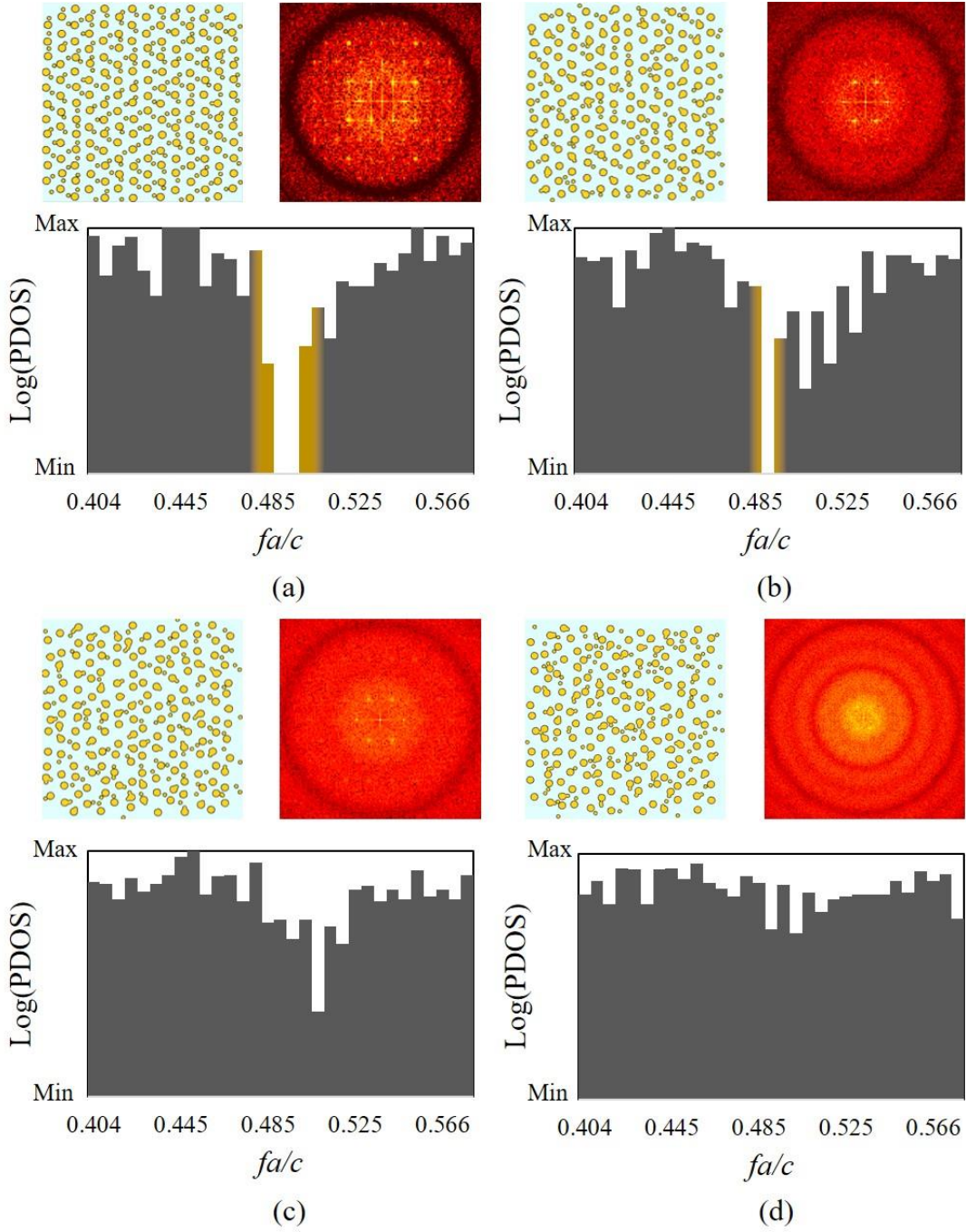

Fig. 4. Superlattices (upper left), Fourier transformations (upper right), and PDOS (bottom) under different degrees of deformation: (a) $R_d = 1\times(L/2-r_i)$. (b) $R_d = 2\times(L/2-r_i)$. (c) $R_d = 3\times(L/2-r_i)$. (d) $R_d = 4\times(L/2-r_i)$.

We further verify valley transport along the interface of photonic crystals under the deformation ($R_d = L/2-r_i$). The topological index difference at the K/K' points lead to unidirectional edge modes. Fig. 5 shows two different interfaces: straight and triangular interfaces. Unidirectional propagation can be observed in periodic structures and deformed structures, as shown in Fig. 5(a, b, c, and d). The circularly polarized excitation is set at one side of the interface waveguide, and OAM modes are excited at valley points, as shown in Fig. 5(e). Probes at the output side are set to obtain the transmission spectra in Fig. 5(f). Maximum and minimum transmission points in deformed structures are marked as circles and squares, where corresponding field distributions are shown in Fig. 5(b and d) and Fig. 6, respectively. Transmission bandwidths of straight interfaces are wider than bandwidths of triangular interfaces because straight interfaces support propagation along the y-axis, while triangular interfaces support propagation towards both the y-axis and at an angle of 60-degrees. That means waves in more directions needs to be forbidden in the case of triangular interface than waves in the case of straight interface. Thus, it is natural to observe a narrower bandwidth when we bend a straight interface into a triangular interface. When deformations happen, the bandwidth becomes further narrower as shown in Fig. 4. Besides the change of bandwidth, transmission along interfaces of deformed structures varies because the excitation sources are no longer at the strict center of the deformed hexagonal unit cell, which causes inevitable mismatch between the source and the intended spin mode. Thus, the transmission efficiency is almost perfect at some frequencies, where ideal unidirectional transport can be observed clearly, as shown in Fig. 5(b and d). However, at some frequencies within the bandgap, transmission efficiency is low, as marked in Fig. 5(f).

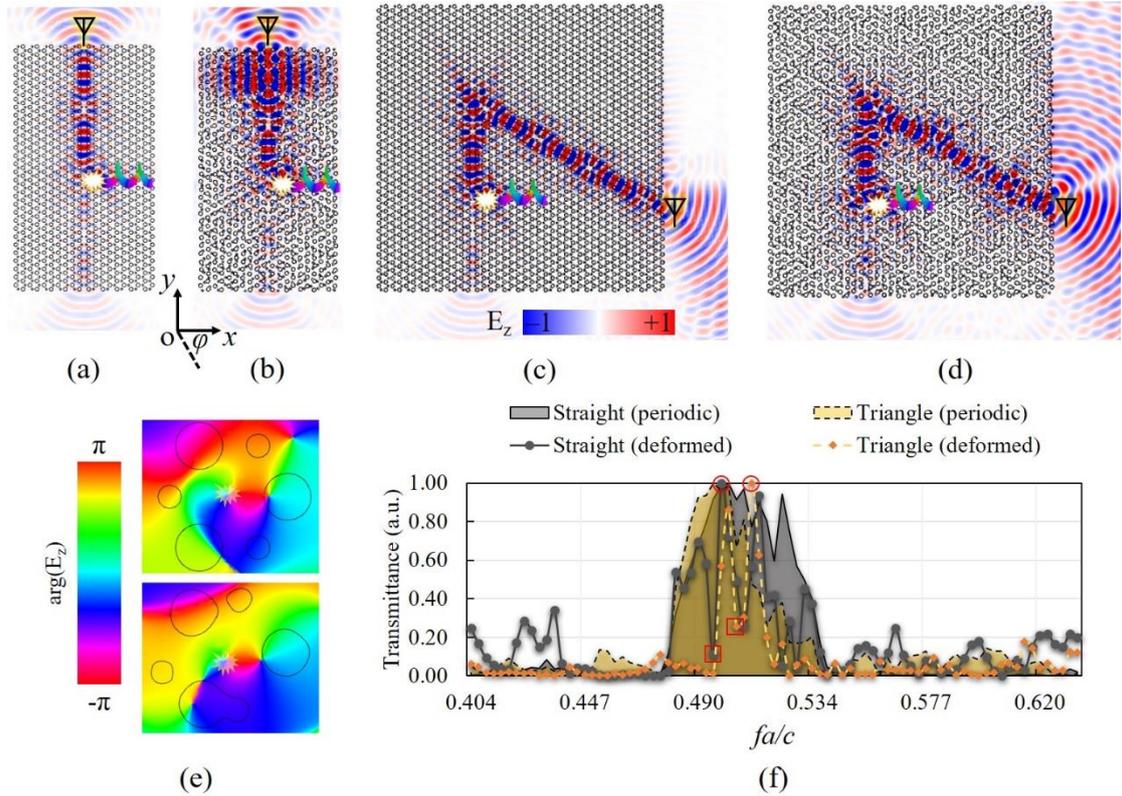

Fig. 5. Topological transports excited by chiral sources ($M_{//} = M_x + jM_y$). Straight interface: (a) periodic structure, (b) deformed structure. Triangular interface: (c) periodic structure, (d) deformed structure. (e) The chiral point source is utilized to excite OAM modes at valley points in a hexagonal unit cell. (f) Normalized transmission spectra where maximum and minimum transmission points in deformed structures are mark as circles and squares respectively.

Electric field distributions at these minimum transmission points are shown in Fig. 6. The point source is set at the same position, and we can observe the edge modes along the interface instead of the unidirectional propagation shown in Fig. 5(b and d), the maximum transmission points. Radiation happens at both ends of the interfaces clearly, which is a typical property of valley transport [24]. It means that in a random deformed structure, it is difficult to excite pure unidirectional transport within the whole bandgap due to the mismatch between the circularly polarized source and eigenmodes of a random deformed hexagon at valley points. Although the deformed bulk still reserves the statistical topological bandgap, eigenmodes of any particular deformed hexagon are unpredictable. The excitation can be divided into two orthogonal spin modes, propagating towards two opposite directions along the interface and at the end of the interface, radiation occurs.

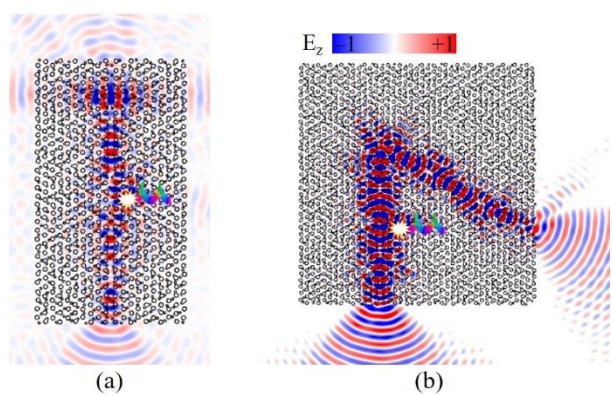

Fig. 6 Electric field distribution of edge modes at minimum transmission points within the photonic bandgap: (a) straight interface, (b) triangular interface.

We further utilized a 3D printing technique to design and fabricate a sandwich-like sample to prove the existence of a topological edge mode in a system with random deformations. Fig. 7(a) shows eigenmodes at valley points of an original sandwich-like unit cell with periodic boundary condition. Fig. 7(b) shows the structure undergoing random deformations, and the inset shows the excitation setup. Two fabricated samples with straight and triangular interfaces are shown in Fig. 7(c and d) respectively, where positions of two probes connected to a vector network analyzer (Keysight PNA N5224A) are marked. In Fig. 7(e), simulated and measured transmissions are compared. The simulation setup is exactly the same as in the experiments, where linearly polarized probes are set at the center of the interface to measure the transmissions. We also removed the dielectric structure in the middle layer and measured background transmission of the parallel-plate waveguide as a benchmark. The background transmission is mainly brought by parallel-plate modes, which are difficult to eliminate [42]. The background transmission is much higher in the low frequency band and becomes almost neglectable within the topological bandgap. After assembling the sandwich-like waveguides, the transmission within the topological bandgap increases dramatically, and due to the existence of bulk modes out of the topological bandgap, transmission out of the bandgap is also enhanced to some extent, making it difficult to observe the topological transport. Similar to previous works [22,26,40,43–45], we can further distinguish topological transmission from bulk transmission by minimum transmission points on both sides of topological bandgap.

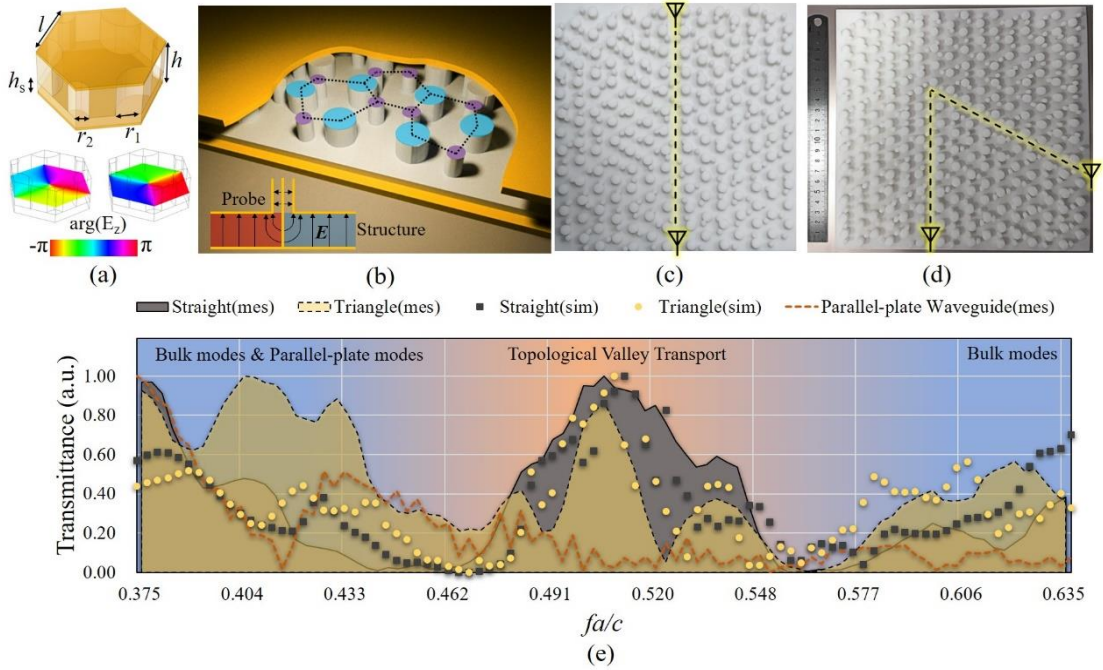

Fig. 7 (a) The ordered metal-dielectric-metal unit cell with periodic boundary conditions working in the microwave band and corresponding phase distributions of $E_z$ of two OAM modes at valley points. Geometric parameters: the total thickness of dielectric structure is $h = 8$ mm, and the thickness of the dielectric plate is $h_s = 2$ mm, radii of the rods are $r_1 = 4$ mm and $r_2 = 2$ mm, the side length $l = 10$ mm. (b) The sandwich-like structure under random deformation $R_d = L/2 – r_i$, the inset shows the setup of probes at the interface of two topological structures. Fabricated samples with: (c) straight interface, (d) triangular interface. (e) Simulated and measured transmission, where results of two samples are normalized. The measured transmission property of a parallel-plate waveguide with 8 mm gap, exactly the same as the structure height, is also shown as a benchmark.

Note that we set probes perpendicular to the metal plates to excite and detect the TM polarized mode. The ideal method to excite topological edge modes is to use chiral waves at the center of a hexagonal unit cell [20,37,46], but a localized source at the interface can also excite the topological edge mode as well [22,47–49]. As mentioned before, the eigenmode at valley points of any deformed hexagonal cell cannot be a pure OAM mode. Even if we keep a hexagonal lattice undeformed, the structure still lacks strict periodic boundary conditions to ensure an ideal OAM mode. Therefore, we have no choice but to place a linearly polarized probe at the center of the interface to excite the edge transport along the interface. We could not obtain the exact transmission efficiency of the deformed structure due to the lack of transition designs between probe and the interface. However, by comparing transmission spectra of the straight interface and the triangular interface as shown in Fig. 7(e), we found that transmission peaks of straight and triangular interfaces are almost the same, demonstrating that the topological edge mode is immune to the sharp defect of the interface. The bandwidth of the straight interface is wider than the bandwidth of the triangular interface, similar to the discussion about Fig. 5(f).

In conclusion, we investigate photonic topological insulators under long-range deformations, where the transition from the ordered to the amorphous structure is discussed. As long as statistical features, such as the bandgap shown in PDOS and peaks of the Fourier spectra are recognizable, the bulk can maintain its topological properties. We further designed and conducted experiments based on a 3D printing technique. Excitation of the deformed system is a challenging task. Traditional chiral wave coupling methods are no longer valid due the lack of strict periodicity. Even in a slightly deformed lattice, mismatch between a chiral source and the intrinsic modes of a particular unit cell is obvious. Thus, a linearly polarized source at the center of interface acts as a more robust method. Transmissions of straight and triangular interfaces were simulated and measured to prove the robustness of topological transport in the deformed structure. This study extends investigations of topological transport from local disorder to long-range deformations of the whole lattice.

**Acknowledgements**
This work was supported by National Natural Science Foundation of China (Nos. 61771127 and 61427801), Scientific Research Foundation of Graduate School of Southeast University (No. YBJJ1814), and Postgraduate Research & Practice Innovation Program of Jiangsu Province (No. KYCX18_0098) as well as AFOSR contract FA9550-16-1-0093.